\documentclass[twocolumn]{emulateapj}
\usepackage{color, natbib}
\usepackage{hyperref}
\shorttitle{Rigel Variability}
\shortauthors{Moravveji, Guinan et al.}

\newcommand{\most}{\textit{MOST}}

\newcommand{\teff}{T_{\mbox{\scriptsize eff}}}

\newcommand{\epsnuc}{$\epsilon_{\mbox{\scriptsize nuc}}$}
\newcommand{\eps}{$\epsilon$-mechanism}
\newcommand{\kap}{$\kappa$-mechanism}

\newcommand{\gov}{g$_{\mbox{\scriptsize ov}}$}
\newcommand{\gst}{g$_{\mbox{\scriptsize st}}$}
\newcommand{\thr}{\tau_{\mbox{\scriptsize th}} }
\newcommand{\dyn}{\tau_{\mbox{\scriptsize dyn}}}
\begin{document}
% =================================================================================================================
% ================================================= HEADING =======================================================
% =================================================================================================================
\title{Asteroseismology of the Nearby SN-II Progenitor Rigel \\
        Part \textsc{II}. $\epsilon-$Mechanism Triggering Gravity-Mode Pulsations?}

\author{Ehsan Moravveji\altaffilmark{1,3}}
\altaffiltext{1}{Department of Physics, Institute for Advanced Studies in Basic Sciences (IASBS), Zanjan 45137-66731, Iran}
\email{moravveji@iasbs.ac.ir}

\author{Andres Moya\altaffilmark{2}}
\altaffiltext{2}{Departamento de Astrof\'{\i}sica, Centro de Astrobiolog\'ia (INTA-CSIC), PO BOX 78, 28691 Villanueva de la Ca\~nada, Madrid, Spain}

\author{Edward F. Guinan\altaffilmark{3}}
\altaffiltext{3}{Department of Astronomy, Villanova University, 800 Lancaster Ave, Villanova PA, USA}

% =================================================================================================================
% ================================================= ABSTRACT ======================================================
% =================================================================================================================
\begin{abstract}
The cores of luminous B and A-type (BA) supergiant stars are the seeds of later core collapse supernovae. 
Thus, constraining the near-core conditions in this class of stars can place tighter constraints on the size, mass and chemical 
composition of supernova remnants. 
Asteroseismology of these massive stars is one possible approach into such investigations.
Recently, Moravveji et al. (2012, hereafter Paper \textsc{I}) extracted 19 significant frequencies from a 6-year radial velocity 
monitoring or Rigel ($\beta$ Ori, B8 Ia).
The periods they determined broadly range from 1.22 to 74.74 days.
Based on our differentially rotating stellar structure and evolution model, Rigel, at itÕs current evolutionary state, is 
undergoing core He burning and shell H burning.
Linear fully non-adiabatic non-radial stability analyses result in the excitation of a dense spectrum of non-radial gravity-dominated mixed modes. 
The fundamental radial mode ($\ell=0$) and its overtones are all stable. 
When the hydrogen burning shell is located even partially in the radiative zone, a favorable condition for destabilization of
g-modes through the so-called \eps ~becomes viable.
Only those g-modes that have high relative amplitudes in the hydrogen burning (radiative) zone can survive the strong radiative damping. 
From the entire observed range of variability periods of Rigel (found in Paper \textsc{I}), and based on our model, only those modes 
with periods ranging between 21 to 127 days can be theoretically explained by the \eps. 
The origin of the short-period variations (found in Paper \textsc{I}) still remain unexplained.
Because Rigel is similar to other massive BA supergiants, we believe that the \eps ~may be able to explain the long-period variations
in $\alpha$ Cygni class of pulsating stars.
\end{abstract}
% =================================================================================================================
% ================================================= SECTION =======================================================
% =================================================================================================================
\section{Introduction}\label{s:intro}
After the core hydrogen depletion, massive stars ($M\gtrsim15 M_\odot$) enter the BA supergiant phase of stellar evolution.
During this epoch of evolution, they exhibit periodic and/or quasi-periodic low amplitude flux microvariability in addition to  variations in radial velocity 
and equivalent width of spectral lines 
\citep{waelkens-1998-01, kaufer-1996-01, kaufer-1997-01,bresolin-2004-01}.
Spectroscopic analysis of \citet{lefever-2007-01} clarified that blue supergiants (BSGs) pulsate in the gravity mode.
Recent studies \citep{degroote-2010-01, noels-2010-01, miglio-2009-01} show that asteroseismology of massive BSG stars
can probe the extent of convective core overshooting, based on the observed and predicted regularity in period spacings.

On the other hand, there are various theoretical explanations for the observed oscillations in BSGs.
The photometeric study of \citet{saio-2006-01} on the B2 Ib/II BSG HD 163899 (which is most likely a less-evolved less-massive analogue of Rigel)
showed that the \kap ~can excite a rich spectrum of pressure (p) and gravity (g) modes in the post-terminal-age-main-sequence stars. 
More importantly, they showed that only those modes that arrive with an appropriate phase to the base of intermediate convective zone (ICZ) 
can be reflected back to the surface, and hence be observed.
\citet{gautschy-2009} searched for the origin of long-period variabilities in the prototype 
$\alpha$ Cyg. Interestingly his Figure 5 shows a gap for stellar models with $3.95\lesssim\log\teff\lesssim4.15$ where 
no instability is predicted; Rigel lies in this gap.
According to \citet{saio-2011-01}, Rigel should be unstable against non-radial convective g$^-$ modes.
Radial and non-radial \textit{strange} mode is proposed as another mechanism to induce instability and interplay with mass loss efficiencies 
in those massive stars with $\log(L/M)\gtrsim4\log(L_\odot/M_\odot)$ 
\citep{glatzel-1993-01, saio-1998-01, glatzel-1999-01, dziembowski-2005-01, aerts-2010-02, saio-2011-01}.
This requirement is also marginally fulfilled by Rigel.
\citet{godart-2009-01} investigated the destructive impact of core overshooting and mass loss during MS on the extent of ICZ, 
and showed that models with wider ICZs are more likely to destabilize stellar oscillations. 
%
% ------------------------------- Table -------------------------------
  \begin{deluxetable*}{lcccccccccc}
  \tablecaption{Observed Rigel properties and \texttt{MESA} Input Models. \label{t:mesa} }
%  \tablewidth{6cm}
  \tabletypesize{\scriptsize}
  \tablecolumns{11}
  \tablehead{\colhead{ } & \colhead{$M_{\mbox{\scriptsize ZAMS}}$} & \colhead{$M_{\mbox{\scriptsize end}}$}\tablenotemark{a} & 
  \colhead{$\teff$} & \colhead{$\log(L/L_\odot)$} & \colhead{$\log g$} & \colhead{$R$} & 
  \colhead{$Y_{s}$} & \colhead{(N/C)$_{\mbox{\scriptsize s}}$} & \colhead{(N/O)$_{\mbox{\scriptsize s}}$} & \colhead{$v$} \\
  \colhead{ } & \colhead{[$M_\odot$]} & \colhead{[$M_\odot$]} & \colhead{[K]} & \colhead{} & \colhead{[cm s$^{-2}$]} & 
  \colhead{[$R_\odot$]} & \colhead{} & \colhead{} & \colhead{} & \colhead{[km s$^{-1}$]} }
  \startdata
  Literature & \nodata & 21$\pm$3 & 12\,100$\pm$150 & $5.07^{+0.10}_{-0.18}$ & 1.75$\pm$0.10 & 78.9$\pm$7.4 & 0.32$\pm$0.04 & 1.74$\pm$0.60 & 
  0.52$\pm$0.13 & 25$\pm3\tablenotemark{b}$  \\
  Reference & (1) & (1) & (2) & (3) & (2) & (4,\, 5) & (2) & (2) & (2) & (2,\, 6)  \\ 
  Modeled & 19.0 & 18.1 & 12\,058 & 5.09 & 1.88 & 80.5 & 0.32 & 1.75 & 0.50 &76  % 
  \enddata
  \tablecomments{(a) \citet{przybilla-2010-01} suggest $M=21\pm3M_\odot$ by fitting to Geneva rotating models. (b) this is the measured $v\,\sin i$.}
  \tablerefs{(1). This work, (2). \citet{przybilla-2010-01} (3). Paper \textsc{I}, (4). \citet{vanleeuwen-2007-01},
                   (5). \citet{aufdenberg-2008-01}, (6). \citet{simon-diaz-2010-01}}
  \end{deluxetable*}
% ------------------------------- Table -------------------------------
%

Once the MS evolution of massive stars ends due to the H depletion in the core, the CNO fusion which takes place in the 
hydrogen burning shell (hereafter HBSh) is the main source of energy.
Further contraction and heating of the helium-rich core initiates core He burning (CHeB) when the star is a late B supergiant.
While the triple-alpha (3$\alpha$) and CNO reactions have mild $\rho-$dependence, they are both sensitive functions of temperature.
In the zeroth order, the energy generation rate can be expressed as 
$\epsilon_{\mbox{\scriptsize nuc}} \propto \rho^{\mu}T^{\nu}$ where
\begin{mathletters}
\begin{eqnarray}
&& \nu = \left( \frac{\partial \ln \epsilon_{\mbox{\scriptsize nuc}}}{\partial \ln T}\right)_\rho \equiv \epsilon_T, \label{e:eps_T}\\
&& \mu =  \left( \frac{\partial \ln \epsilon_{\mbox{\scriptsize nuc}}}{\partial \ln \rho}\right)_T \equiv \epsilon_\rho. \label{e:eps_rho}
\end{eqnarray}
\end{mathletters}
Regarding the 3$\alpha$ reaction, $\epsilon_\rho = $ 2, and $\epsilon_T=$ 40 to 19 for $T_8=1$ to 2 in addition to an extra $Y^3$ dependence.
Similarly, for the CNO burning network $\epsilon_\rho = $ 1, and $\epsilon_T=$ 13 to 17 at $T_6=$ 50 to 25. 
$T_6$ and $T_8$ denote temperature in million and hundred million Kelvin degrees, respectively, and $Y$ is the fractional He abundance.
For more detailed discussion see \cite{clayton-1968}, 
\citet{weiss-2004-01}, 
\citet{maeder-2009-book} and \citet{wiescher-2010-01}. 

Several authors have shown that during the evolution of low mass stars \citep{noels-1974-01, noels-1976-01, sonoi-2011-01} 
and 
massive stars \citep{shibahashi-1976-01}, there occurs conditions where the \epsnuc ~can inject enough energy into the mechanical energy
required for the propagation of vibrational modes, and overcome strong radiative damping close to the stellar core. 
This is the so called \eps ~\citep{unno-1989}. 
This mechanism can be operative during shell H burning \citep{kawaler-1988-01} or shell He burning \citep{kawaler-1986-01}
phase to successfully explain the observed gravity mode instabilities in hot white dwarf stars. 
The radial fundamental mode of very low mass stars and brown dwarfs can be also unstable during the central deuterium burning phase
\citep{palla-2005-01, rodrigez-2011-01}. 
Most recently \citet{miller-bertolami-2011-01} applied the same mechanism to a subdwarf B star prior to CHeB phase.
In accord with these, the favorable conditions for excitation of internal g-modes can also take place in post-MS massive stars, and lead to 
pulsational instabilities.

We find Rigel to be an excellent testbed to investigate the existence of the \eps ~ to excite oscillatory modes and to study the possible 
asteroseismic potential of slowly pulsating blue supergiant stars. 
In a previous work, \citet[][hereafter Paper \textsc{I}]{moravveji-2012-01} presented the \most ~photometry and radial velocity variations of 
Rigel, and tabulated 19 significant pulsation frequencies from radial velocity variations. 
The periods of observed modes range from 1.21 to 74.74 days. 
Table \ref{t:mesa} summarizes the observed physical properties of Rigel from the literature which are an invaluable input for 
the equilibrium modeling of the star. 
Y$_{\mbox{\scriptsize s}}$, (N/C)$_{\mbox{\scriptsize s}}$, and (N/O)$_{\mbox{\scriptsize s}}$ are surface abundance ratios
of He and CNO products, respectively.
The column marked with $v$ shows the measured $v \sin i$ \citep{przybilla-2010-01,simon-diaz-2010-01}. 
We explain the properties of the model we found to describe Rigel in Section \ref{s:mesa}. 
The pulsation analysis is discussed
in Section \ref{s:seismic} and in Section \ref{s:discuss} we summarize the conditions for the occurrence of the \eps.
The purpose of this paper is to highlight the importance of the \eps ~as a possible destabilizing driving force agains pulsations 
in BSG core-collapse progenitors \citep{murphy-2004-01}.
%
% =================================================================================================================
% ================================================= SECTION =======================================================
% =================================================================================================================
\section{Modeling Rigel}\label{s:mesa}
% ------------------------------- Figure -------------------------------
\begin{figure*}[t!]
\epsscale{1.2} \plotone{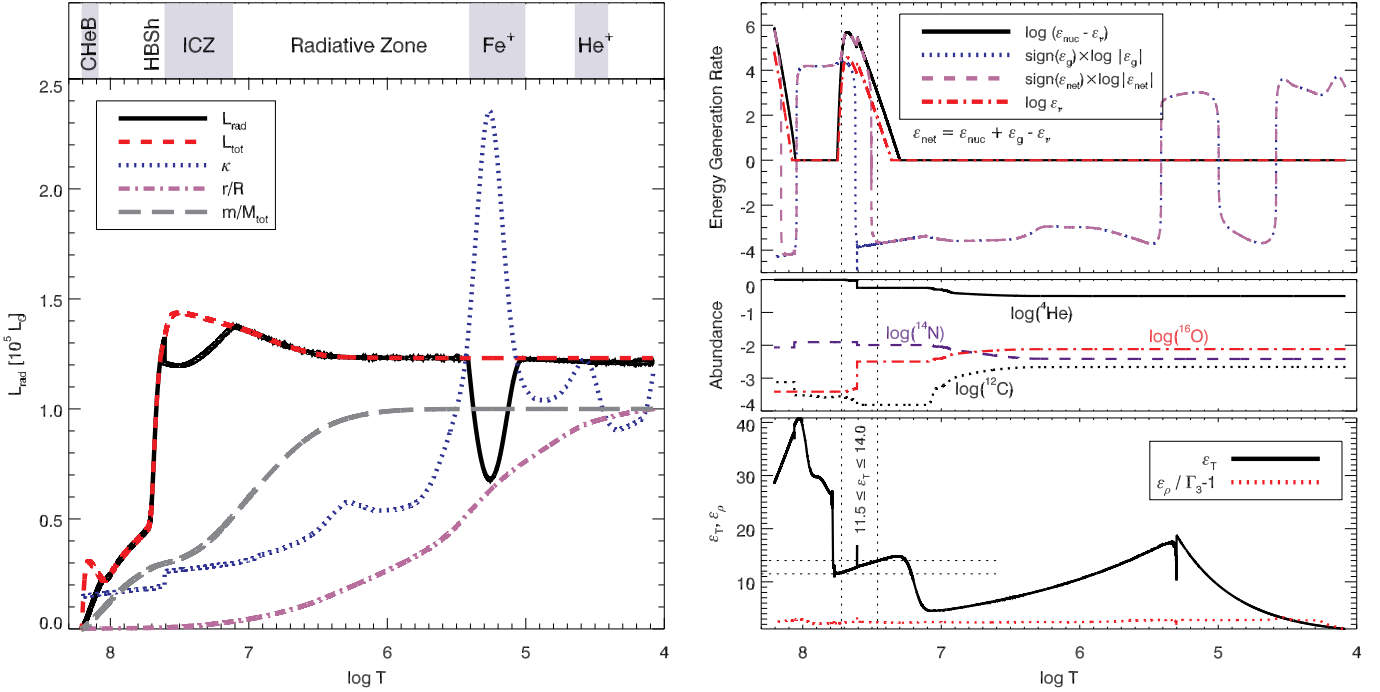}
\caption{Internal structure of the input model. 
On the left, the profile of total luminosity (short dashed line), radiative luminosity (solid line) and opacity (dotted line) along with 
distribution of mass (long dashed line) and radius (dashed-dot line) are plotted against $\log T$.
On the top right panel is the profile of \epsnuc ~(solid line), gravitational energy (dotted line), neutrino energy loss (dashed-doted line) and 
the net energy (dashed line). The right middle panel shows the abundance of He and CNO cycle elements. 
The bottom right panel shows the temperature (Eq. \ref{e:eps_T}) and density (Eq. \ref{e:eps_rho}) derivatives of \epsnuc, respectively.
The vertical lines mark the extent of HBSh, and show the limits of $\epsilon_T$ which is favorable in this zone for \eps.
\label{f:interior}}
\end{figure*}
\subsection{Input \texttt{MESA} Model}\label{ss:mesa}

A model that can represent the current evolutionary stage of Rigel on the Hertzsprung-Russell diagram (HRD) has been
generated with the aid of \href{http://mesa.sourceforge.net/}{\texttt{MESA}} stellar structure and evolution code (version 3723). 
A full description of this instrument is given by \citet{paxton-2011-01}. 
A non-magnetic differentially rotating model with the initial mass $M_{\mbox{\scriptsize ZAMS}}=19 M_\odot$ and
initial solar composition as in \citet{grevesse-1993-01} 
at solar metallicity $Z=0.02$ were generated
with the initial rotation velocity $v_i=200$ km s$^{-1}$ to help mixing He and CNO products to the surface
and reproduce the observed values. 
Surface abundance yields up to the pre-core-collapse stage were compared with \citet{heger-2005-01} to ensure for their agreement.
Note that our choice of initial abundances complies with those given in Tables 1 of \citet{przybilla-2010-01}.
The surface of the model is located at the optical depth $\tau=0.667$ using 
Eddington $T-\tau$ integration scheme \citep{eddington-1926-01}. 
Mixing Length theory of \citet{bohm-vitense-1958-01} is used to treat the convective mixing.
The borders of convective zones are found using the Schwarzschild criteria 
($\nabla_{\mbox{\scriptsize rad}}=\nabla_{\mbox{\scriptsize ad}}$).
The model was evolved with the above mentioned initial values until the post-MS phase and stopped when $\teff=12\,058$ K,
close to the recent measurement.
The final radius of the model proved to have negligible dependence on the choice of the mixing length parameter (which is defined 
as the ratio of the length over which convective cells loose their identity to the local pressure scale height 
$\alpha_{\mbox{\scriptsize MLT}}=L/H_p$).
A value of $\alpha_{\mbox{\scriptsize MLT}}=1.6$ was adopted for the model. 
The overshooting from the top and bottom of convective zones were suppressed. % \citep{godart-2009-01, aerts-2011-01}.
The metalicity-dependent mass loss recipe of \citet{vink-2001-01} and \citet{glebbeek-2009-01} with 
$\eta=0.8$ as in \citet{maeder-2001-01} is used during the evolution across the HRD in all models. 
Calculation of Brunt-V\"{a}is\"{a}l\"{a} frequency is based on Equation 14 in \citet{brassard-1991-01}.
To reach the surface yields of He and CNO as measured by \citet{przybilla-2010-01}, 
we enhanced the efficiency of rotational mixing by a factor of two \cite[$f_c=2$ in Eq. 53 in][]{heger-2000-01}. 

Table \ref{t:mesa} lists the modeled physical parameters of Rigel to compare with observations.
Not all calculated models meet their measured counterparts. 
It is neither easy to find a model that simultaneously matches all observed quantities and nor our goal to present the ``best" model of Rigel. 
Yet, the adopted model closely represents the current observed and physical properties of Rigel, and furthermore yields consistent seismic results.
\subsection{Internal Structure of the Adopted Model}\label{ss:interior}
To comprehend better the complex internal structure of our model, Figure \ref{f:interior} shows profiles of several thermodynamic 
quantities with $\log T$ on the abscissa. 
On the left panel, the solid line shows the profile of radiative luminosity which starts to grow 
from the core due to the CHeB followed by a steep increase around $\log T=7.7$ arising from HBSh; 
this profile deviates from total luminosity (short dashed line) wherever convection assists radiation in energy transport.  
Therefore, the dips in radiative luminosity profile mark the convective zones. 
Most relevant to our study are three radiative zones and three convective zones. 
They have profound impacts on the energy balance of pulsation modes.
The ICZ is situated just above the HBSh at $7.1 \leq \log T \leq 7.6$;
the Iron- and He-bumps of opacity are around $\log T\simeq5.2$ and  $\log T=4.6$, respectively.
There are two radiative zones below and above the ICZ, where the former hosts the peak of energy generated from H burning,
and the latter lies above the ICZ.
There is another radiative zone surrounded by the two opacity bumps.
The long-dashed and dash-dotted lines signify the spatial mass distribution of the model.
Accordingly, around 99\% of the mass is inside 25\% of the star's radius.  

Right panels of Figure \ref{f:interior} show profiles of various quantities related to the energy generation rate. 
On the top panel, the solid line is the run of energy generation rate $\epsilon_{\mbox{\scriptsize nuc}}$ in logarithmic scale. 
The CHeB peak in the core and HBSh peak at $\log T=7.7$ are evident.
The dash-dotted line shows the neutrino cooling rate $\epsilon_\nu$ which is mainly an order of magnitude smaller than 
$\epsilon_{\mbox{\scriptsize nuc}}$ except below HBSh. 
There is an extra source (sink) of energy depending on whether a zone is contracting (expanding). 
This is designated by $\epsilon_g = -T dS/dt$ (dotted line) where $T$ and $dS/dt$ denote temperature and entropy time derivative, respectively.
Note that for the layers undergoing expansion, $\epsilon_g$ is negative. 
The radiative zone above the ICZ is where expansion is dominant.
Finally, the profile of the generated net energy $\epsilon_{\mbox{\scriptsize net}}$ is shown by the dashed line. 
Note that in the non-burning envelope, expansion and contraction are sources and sinks of energy. 
Two vertical dotted lines show the extent in the model where energy is purely generated by CNO cycle.
A fascinating feature in this model is that this shell is situated partially in a radiative zone which is central to our study.

The right middle panel of Figure \ref{f:interior} shows the runs of $^4$He, $^{12}$C, $^{14}$N and $^{16}$O abundances. 
The ratios of $^{12}$C/$^{14}$N$=0.015$
and $^{16}$O/$^{14}$N$=0.30$ are comparable with those from Table 25.3 of \citet{maeder-2009-book} in the HBSh zone.
% Ehsan:  Ratio               Maeder                MESA
%              C/N                   0.025                  0.015
%              O/N                   0.10                    0.300
%              C/C13               3.3                      3.56

The right bottom panel of Figure \ref{f:interior} shows the logarithmic derivative of
energy generation rate $\epsilon_T$ (solid line, Eq. \ref{e:eps_T}) and $\epsilon_\rho$ (dotted line, Eq. \ref{e:eps_rho}) for our model, respectively. 
Consistently, $\epsilon_T$ varies from $\sim$30 to 40 in the Helium rich core, and steeply declines to $\sim$11.5 to 14.0 in the HBSh where CNO is contributing 
to a large fraction of Rigel's luminosity.
This latter range of $\epsilon_T$ is one of the favorable conditions for \eps ~to operate (Section \ref{ss:g-modes}).
A caution must be given to a peak in $\epsilon_T$ at non-burning envelope around $\log T=5.3$ where 
$\epsilon_{\mbox{\scriptsize nuc}}$ is vanishingly small.
This arises from careful numerical differencing followed up in \texttt{MESA} and will have no effect on our modal stability analysis
since the combination $\epsilon_{\mbox{\scriptsize nuc}}$ times $\epsilon_T$ play the major roles.
%
% ------------------------------- Section -------------------------------
\subsection{Propagation Cavities}
\begin{figure}[t]
\epsscale{1.15} \plotone{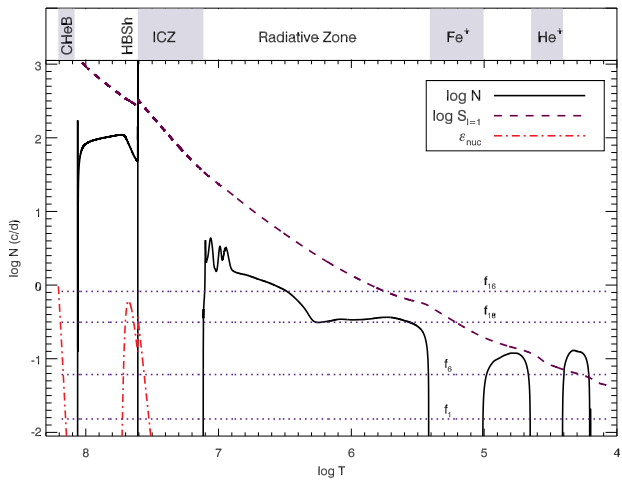}
\caption{Propagation diagram for the model described in Table \ref{t:mesa}. 
The Brunt-V\"{a}is\"{a}l\"{a}  $N$ (solid line) and the Lamb frequency $S_{\ell=1}$ (dashed line) are in logarithmic scale.
The radiative (convective) zones are where $N^2>0$ ($N^2<0$). 
The profile of \epsnuc ~is in an arbitrary unit only to show that the peak of \epsnuc ~at HBSh resides in the innermost radiative zone.
Four selected frequencies from (Table 2 in) Paper \textsc{I} are also shown as horizontal dotted lines to mark the propagation cavity of each mode.
\label{f:brunt}}
\end{figure}

The Brunt-V\"{a}is\"{a}l\"{a} frequency $N^2$ which is defined as \citep{brassard-1991-01}
\begin{equation}\label{e:brunt}
N^2 = \frac{g}{r}\left[ \frac{1}{\Gamma_1}\frac{\partial \ln P}{\partial \ln r} - \frac{\partial \ln \rho}{\partial \ln r} \right].
\end{equation}
shows the extent and boundaries of convective zones ($N^2<0$ where g-modes are evanescent) and the radiative zones 
($N^2>0$ where g-modes are oscillatory).
The complexity of the internal structure of our model is indicated in Figure \ref{f:brunt} which shows the run 
of the square root of $N^2$ (solid line) and Lamb frequency $S_\ell^2= \ell(\ell=1)c_s^2/r^2$ as a function of temperature.
Here, $c_s$ denotes the sound speed in the stellar plasma.
The profile of \epsnuc ~is designated by dot-dashed line in arbitrary units.

The remarkable feature in the model is that the peak and hotter part of HBSh reside in the innermost radiative zone 
and extend slightly to the base of ICZ.
The coincidence of this condition is one of the prerequisites for the \eps ~to operate (Section \ref{ss:g-modes}).
This, however, is not true during the entire evolution of a massive star.
Once the central H burning is ceased, the helium rich core starts to contract. 
As a result the core temperature $T_c$ increases, and the opacity $\kappa$ near the core drops. 
In addition, the non-burning core contributes very poorly to the radiative luminosity $L_r$.
Therefore, the radiative temperature gradient $\nabla_{\mbox{\scriptsize rad}}\propto \kappa P L_r / m T^4$ becomes smaller than
$\nabla_{\mbox{\scriptsize ad}}=(\partial \ln T/ \partial \ln P)_S$ mainly because of a drastic decrease in $L_r/m$. 
Thus, the core becomes radiative.
This marks the starting point of the post-MS evolutionary phase.
The HBSh, which at this evolutionary stage, is the main source of energy, moves away from the core and lies partly in the ICZ 
and partly in the innermost radiative zone.

% ------------------------------- Figure -------------------------------
% ------------------------------- Figure -------------------------------

% =================================================================================================================
% ================================================= SECTION =======================================================
% =================================================================================================================
\section{Seismic Analysis}\label{s:seismic}
The model discussed in Section \ref{s:mesa} is the input to the Granada oscillation Code \citep[hereafter GraCo:][]{moya-2004-01,moya-2008-01}.
The radial and non-radial oscillation spectra are calculated in both the adiabatic and non-adiabatic frames
solving numerically the perturbative equations described in \citet{unno-1989}. 
The set of adiabatic equations provide a first list of eigenfrequencies and eigenfunctions. 
In the non-adiabatic approximation, the eigenfrequencies are corrected (slightly for non-adiabatic effects), and an additional set of eigenfunctions 
related with the energy interchanges are obtained.
These eigenfunctions make it possible to calculate theoretically the work done by the pulsational mode to its
surroundings, during a complete period. 
In addition, we can obtain the relative flux variations and the phase-lag between the flux variations and the corresponding variations in radius.
These last two non-adiabatically derived quantities are not used in the present study, but could be useful once multi-wavelength photometry
of Rigel is carried out. 

In the GraCo, the non-adiabatic calculations can be done to include the convection -
pulsation interaction using the Time Dependent Convection (TDC) theory \citep{dupret-2005-01}. 
This theory improves the ``frozen convection'' (FC) approximation usually implemented in most of the other codes.
But this is not necessary in this study, since near the core of stars like Rigel, where the \eps ~is located, 
convection is highly developed.
As a result, convection can be regarded decoupled from pulsation.
In this case the FC is an appropriate approximation. 
On the other hand, the adiabatic solutions of GraCo have been used as reference 
for the ESTA works \citep[Evolution and Asteroseismic Tools Activities,][]{lebreton-2008-01, moya-2008-02}.

To comply with our list of observed frequencies presented in Paper \textsc{I}, 
we restrict our search for frequencies from less than $10^{-2}$ to 1.3 d$^{-1}$ and for radial and non-radial modes up to $\ell=3$. 
Below the lower frequency limit of $10^{-2}$ d$^{-1}$, convergence is very difficult to achieve.
In this respect, our strategy is different from that of \citet{godart-2009-01} where they employ quasi-adiabatic approach for stability analysis of
the radiative core and full non-adiabatic solution for the envelope. 
This approach is optimized to extract a discrete set of excited modes reflected from the edges of the ICZ in those models where the \kap ~is 
responsible for pulsational instabilities.
We note that while \citet{gautschy-2009} did not find any instabilities for models of Rigel ($3.95\lesssim\log\teff\lesssim4.15$), 
\citet{saio-2011-01} predicted that $\alpha$ Cyg type pulsating stars ought be unstable against convective gravity modes.
In the following, we explain the asteroseismic results for different degrees $\ell$.

% ------------------------------- Figure -------------------------------
\begin{figure*}[t!]
\epsscale{1.1} \plottwo{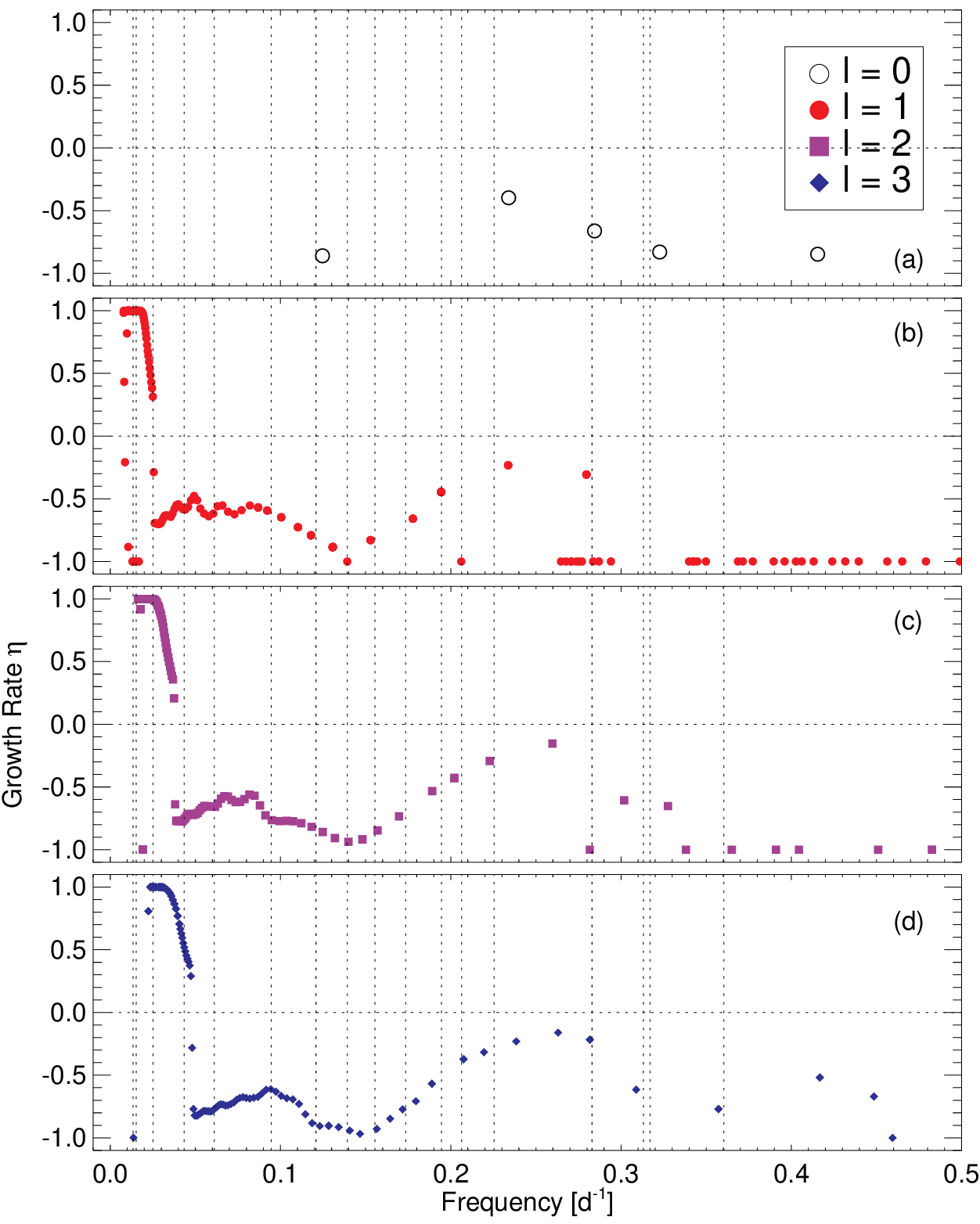}{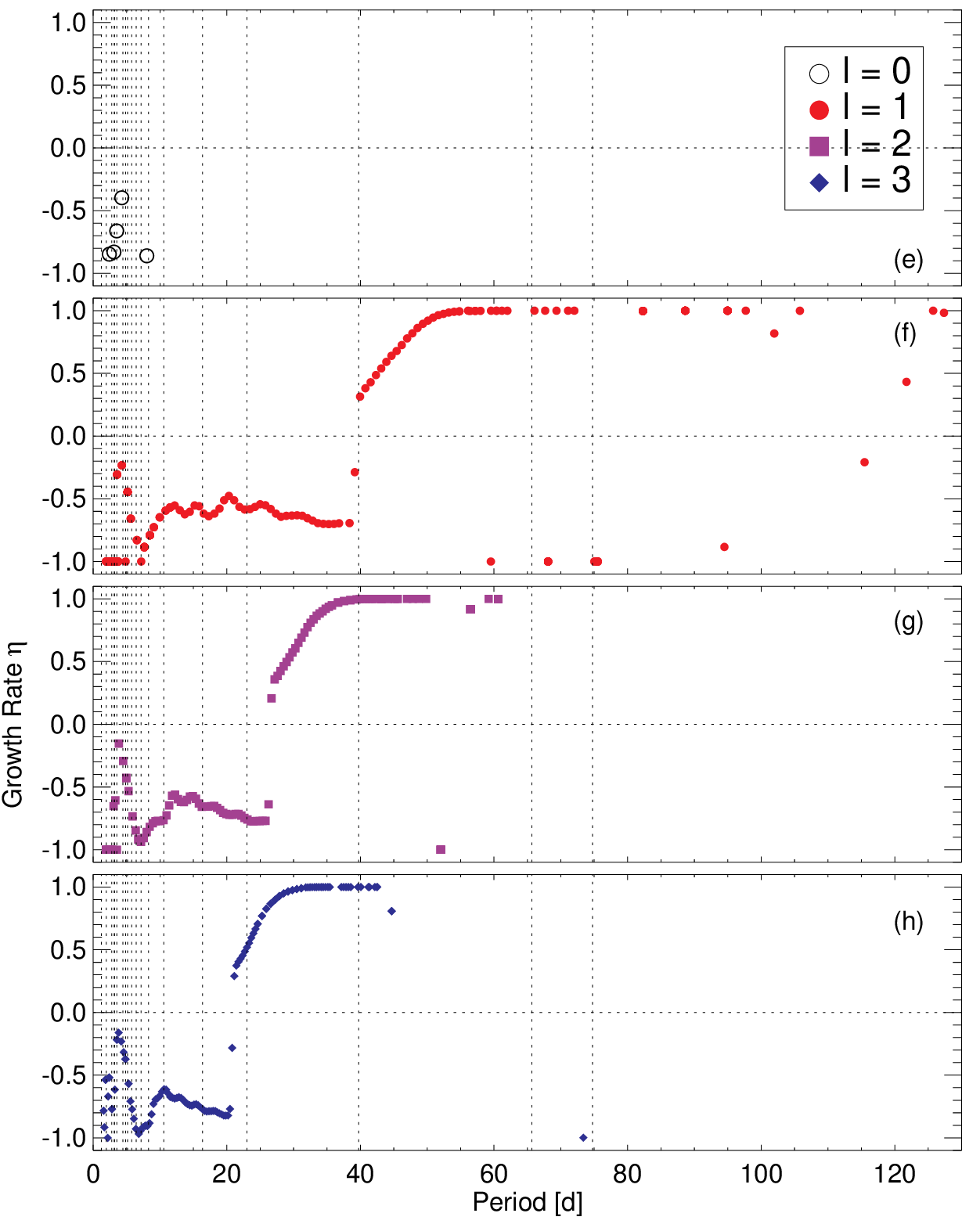}
\caption{ Left. Variation of growth rate $\eta$ with mode frequency for $0 \leq \ell \leq 3$. 
All radial p-modes (empty circles) are stable.
The low-frequency dipole (filled circles), quadrupole (filled squares) and octupole (filled diamonds) modes are 
found excited by the \eps ~($\eta>0$). 
The period of these excited modes vary from about 127 days down to 21 days.
There is a peak in all stable non-radial modes around 0.26 d$^{-1}$ where a tendency towards overstability appears regardless of degree. 
The vertical dotted lines mark the observed frequencies from Paper \textsc{I}.
Right. Similar to the left, but with the periods on the abscissa. 
With the increase in degree $\ell$, the transition from overstable to stable modes shifts towards shorter periods.
\label{f:eta}}
\end{figure*}
% ------------------------------- Figure -------------------------------

\subsection{Energy Balance}\label{ss:energy}
The temporal variation of perturbed quantities in GraCo are treated as $\exp(-i\sigma t)$ where $\sigma_R$ and $\sigma_I$ are the
real and imaginary parts of the complex eigenfrequency $\sigma=\sigma_R+i\sigma_I$,  respectively.
The imaginary part is essentially obtained as a solution of the system of differential equations,
however, it is also related to the so-called cumulative work $W$.
\begin{equation}\label{e:sigma_I}
\sigma_I = - \frac{\sigma_R}{4\pi}\frac{W}{E_k},\qquad 
E_k = \frac{\sigma_R^2}{2} \int_0^M |\xi(r)|^2\, dm.\\
\end{equation}
where $E_k$ is the kinetic energy.
According to \citet{unno-1989} (their Section 26), the work integral accumulates the contribution from perturbation to 
radiative flux $W_F$, convective flux $W_C$, and energy generation rate
$W_\epsilon$, say 
\begin{equation}\label{e:work-sum}
W = W_F + W_C + W_\epsilon,
\end{equation}
The net contribution of  \epsnuc ~to the work integral of each mode is
\begin{equation}\label{e:work}
W_\epsilon = \frac{\pi}{\sigma_R} \int_0^M \epsilon_{\mbox{\scriptsize nuc}} \left( \epsilon_T + \frac{\epsilon_\rho}{\Gamma_3-1} \right) 
       \left( \frac{\delta T}{T} \right)^2\, dm.
\end{equation}
where $\epsilon_T$ and $\epsilon_\rho$ are as in Eq. \ref{e:eps_T} and \ref{e:eps_rho}; $\Gamma_3-1=(\partial \ln T / \partial \ln \rho)_S$ is an
adiabatic exponent and $\delta T$ is the Lagrangian temperature perturbation. 
As the bottom right panel in Figure \ref{f:interior} shows, $\epsilon_T$ and $\epsilon_\rho / (\Gamma_3-1)$ are both positive quantities; 
as a result, the integrand in Eq. \ref{e:work} is greater than zero and always has a tendency towards mode excitation given
high enough values for $\epsilon_T\gtrsim$11 \citep{noels-1974-01, noels-1976-01, sonoi-2011-01}.

We define the $e-$folding time $\tau_e=\sigma_I^{-1}$, and the growth rate $\eta=W / |W|$ where $|W|$ is an absolute value of the work.
With these, growth rate lies in the range $-1.0 \leq \eta \leq 1.0$.
Therefore, the stability of non-radial modes is determined by the sign of growth rate where positive (negative) values 
denote overstability (stability). 

Since \epsnuc ~is practically (but not necessarily numerically) zero in the envelope of our model, the rise in the profile of $\epsilon_T$ 
in (bottom right panel of Figure \ref{f:interior}) has no impact on the evaluation of $W_\epsilon$ due to the multiplication of \epsnuc ~by $\epsilon_T$ (see Eq. \ref{e:work}).

\subsection{Frequency Domain of Excited Gravity Modes}\label{ss:growth}
In evolved stars with alternating radiative and convective zones and large density contrast between the core and the envelope, 
pressure (p) and gravity (g) modes appear in mixed character, where for each, there is a 
contribution from the other. 
Consequently, the modes are identified as p- or g-modes according to the more dominant effect one might have over the 
other \citep{osaki-1975-01}.
In our case, adiabatic eigenfunctions are high-order gravity-dominated mixed modes with large number of nodes in both pressure and gravity cavities.
The huge number of nodes in the \eps ~trapping zone is met by the grid of mesh points which is fully in agreement with the asymptotic approach
\citep{tassoul-1980-01, smeyers-2007-01}.

% ------------------------------- Figure -------------------------------
\begin{figure*}[t]
\epsscale{1.1} \plottwo{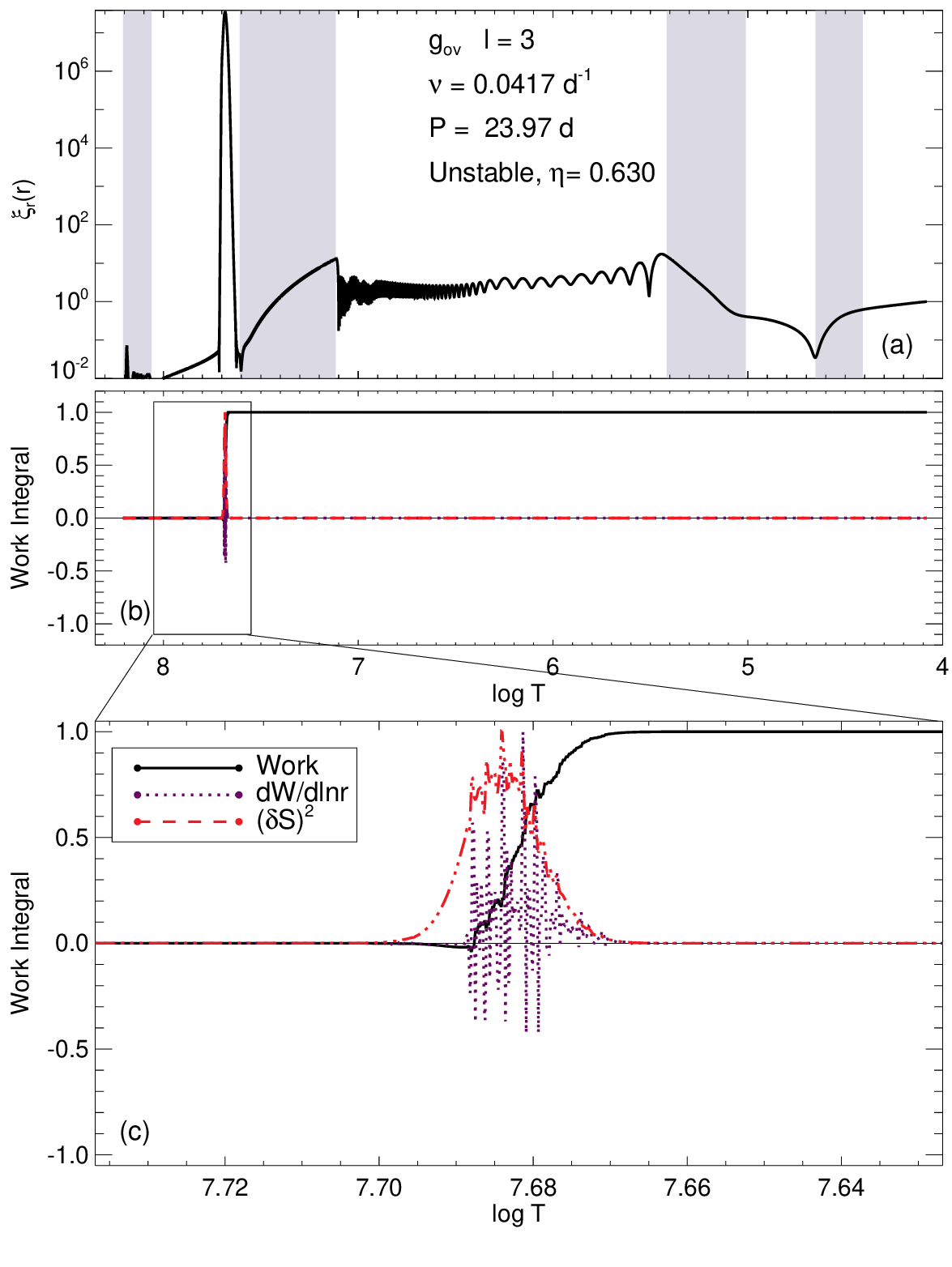}{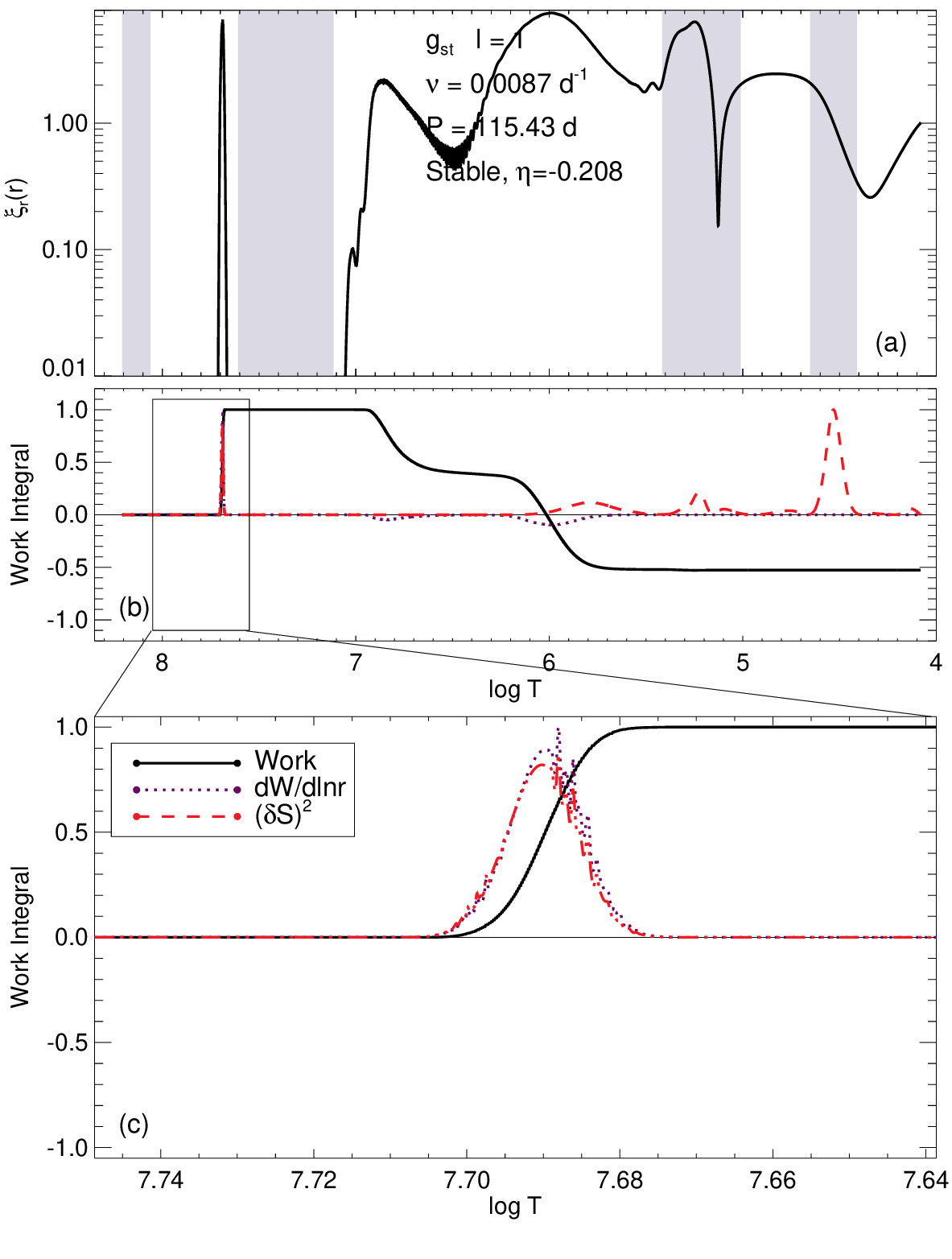}
\caption{An example of two g-modes, \gov ~(left) and \gst ~(right). 
Concise mode information is also included: $\ell, \, \nu, \, P$ and $\eta$ are degree, frequency in d$^{-1}$, period in days and growth rate, respectively.
The modulus of the radial displacements $\xi_r(r)$ in panels (a) are normalized to unity at the surface. 
Though the \eps ~in both modes is the main driver of g-mode pulsation, the latter is stabilized by radiative damping above ICZ. 
The excited mode \gov ~has a large amplitude in the radiative HBSh and above ICZ so that radiative damping is overcome. 
The middle panels (b) show the work integral $W$ (solid black line), work derivative $dW/d\ln r$ (dotted red line) and square of entropy perturbation 
$(\delta S)^2$ (dashed blue line) all normalized to unity. 
The bottom panels (c) are zoomed-in views of the same quantities as in panels (b) in the HBSh.
 \label{f:g-modes}}
\end{figure*}
% ------------------------------- Figure -------------------------------

To study the effect of energy balance, Figure \ref{f:eta} shows the variation of growth rate $\eta$ as a function of mode frequency 
(left panels) and period (right panels) for different spherical harmonic degrees $\ell$. 
Because models of evolved stars have highly non-adiabatic envelopes, the numerical convergence is not straightforward in all cases, so
we do not assume that all eigensolutions are covered;
however, the solution space is well sampled so that there is a clear trend in stability behavior of non-radial modes in Figure \ref{f:eta}. 

For non-radial modes (panels b, c and d in Figure \ref{f:eta}) and at the lower frequency range, the dense spectrum of eigenmodes are
 excited by the \eps ~(discussed in more detail in Section \ref{ss:g-modes}) and the overstability declines 
steeply by frequency with an abrupt transition from unstable to stable modes.
At the high frequency end, all modes are strongly damped $\eta\approx-1.0$. 
In between these two extremes there exist a broad peak in the stable nonradial modes occurring at $\sim$0.26 d$^{-1}$.  
This shows that there is a tendency towards instabilities at higher frequencies whose driving force is different from the low-frequency modes.
As it is discussed in more detail in Section \ref{ss:kappa}, the inefficiency of the \kap ~operating around the He bump of opacity is 
responsible for their stability.
Very clearly, there are many observed frequencies (marked by vertical dotted lines) that have no predicted unstable counterparts.

Panels (e) to (h) in Figure \ref{f:eta} present the profile of the growth rate with the corresponding period of the modes.
The excited modes are more distinct in this representation.
It is clear that four observed modes lie in the predicted domain of excited g-modes destabilized by the \eps.

An intriguing question would be the dependence of the frequency domain of overstable modes on the input physics of the
equilibrium model.
In this way, the low-frequency stable modes could be helped into overstability.
Such physical ingredients could be the increased opacity at Carbon bump ($\log T\approx 6.3$), efficiency of rotationally induced mixing, 
overshooting from the borders of the ICZ and the mass loss rate.
We intend to investigate this in the upcoming studies.

\subsection{Radial Modes}
The Fundamental radial mode ($p_1$ having a period of 8.02 d) and it's overtones in the prescribed scanning frequency range are very stable. 
This stability takes place around the Helium partial ionization zone close to the surface of the star.
A similar result was achieved by \citet{gautschy-2009} which is in very good agreement with the spectroscopic study of 
\citet{lefever-2007-01} who placed their sample of 
BSGs on the instability strip of various classes of pulsating stars and verified that these supergiants are unstable against
g-modes \citep[see also][]{saio-2011-01}.

%
% ------------------------------- Section ------------------------------
\subsection{Excitation of Nonradial Gravity Modes by the \boldmath{ $\epsilon$}-Mechanism}\label{ss:g-modes}
%% ------------------------------- Figure -------------------------------
%\begin{figure}
%\epsscale{1.15}\plotone{ts}
%\caption{The comparison between the ratio of the thermal time scale to the dynamical time scale ($\thr/\dyn$, solid line) 
%versus the dimensionless frequency $\omega=\sigma_R\dyn$, where $\dyn=19.63$ d for the prescribed model.
%The shortest and longest observed frequencies of Rigel, namely $\sigma_R=0.01338$ d$^{-1}$ (dashed line) and 
%$\sigma_R=0.82026$ d$^{-1}$ (dash-dotted line) are taken from Table 2 in \citet{moravveji-2012-01}.
%These two modes are also shown in Figure \ref{f:brunt}.
%\label{f:ts}}
%\end{figure}
%% ------------------------------- Figure -------------------------------

From Eq. 22.1 in \citet{unno-1989} for the linearized equation of energy conservation we have
\begin{eqnarray}\label{e:non-adiab}
i\omega \,  && \left(\frac{\thr}{\dyn}\right) \, \left(\frac{\delta S}{C_p}\right) = 
\frac{4\pi r^3 \rho}{L_r} \left( \delta\epsilon_{\mbox{\scriptsize nuc}} - \frac{d\,\delta L_r}{dM_r}\right) + \nonumber\\ 
&&\frac{\ell(\ell+1)}{d\ln T/d\ln r}\frac{T'}{T} + \ell(\ell+1)\frac{\xi_h}{r}\frac{4\pi r^3 \rho}{L_r}\frac{dL_r}{dM_r},
\end{eqnarray}
where $\omega=\sigma_R\,\dyn$ is the dimensionless eigenfrequency, $\delta S$ is the Lagrangian entropy perturbation,
$\xi_h$ is the tangential eigendisplacement, 
$T'$ is the Eulerian temperature perturbation, and all other symbols have their common meaning.
In the dense interior of the model, the ratio of the thermal time scale $\thr\propto M_r C_p T/L_r$ to the dynamical time scale
$\dyn=(R^3/GM)^{1/2}$ is large.
Even for small $\omega$, the right hand side of Eq. \ref{e:non-adiab} has to be large enough to overcome the high adiabaticity in the interior.
This expectation is fulfilled for those modes which are trapped in the HBSh, for which the perturbation of the nuclear energy generation 
$\delta\epsilon_{\mbox{\scriptsize nuc}}$ is high enough, and other non-adiabatic terms can be overwhelmed.

To illustrate how the \eps ~is capable of exciting g-modes, Figure \ref{f:g-modes} shows two selected solutions, 
an octupole \gov ~(left) and a dipole \gst ~(right).
The former is overstable $\eta = +0.630$ while the latter is stable $\eta=-0.208$. 
The radial component of the eingendisplacement $\xi_r(r)$ which is shown on top panels (a) is set to unity at the surface $\xi_r(R)=1.0$. 
The remarkable feature of \gov ~is its relatively high amplitude in the radiative zone below the ICZ. 
There is a phase lag between the real and imaginary parts of $\xi_r(r)$ which accounts for the huge peak in the modulus of $\xi_r(r)$ as
seen in Figure \ref{f:g-modes}.a.
The energy provided by the \eps ~when combined with large amplitude of \gov ~in HBSh drives the mode and helps it overcome the 
radiative damping.
In panel (b), the run of energy-related quantities are plotted in arbitrary units. 
The abscissa is $\log T$. 
The solid line is the cumulative work integral as in Eq. \ref{e:work-sum}, the dotted line is the work derivative 
$dW/d\ln r$ and the dashed line is the square of the Lagrangian entropy perturbation $\delta S$ 
\begin{equation}\label{e:deltaS}
\delta S = C_p \left( \frac{\delta T}{T} - \nabla_{\mbox{\scriptsize ad}} \frac{\delta p}{p}  \right).
\end{equation}

After a sharp rise in the HBSh, the profile of $W$ flattens until it reaches the surface.
Above the base of the ICZ, the work integral and its derivative are constant, and there is no heat exchange in the envelope, so there $\delta S=0$, 
for this specific solution.
Panel (c) shows a zoomed-in view of the HBSh capable of exciting g-modes. Clearly, the heat exchange and work derivative are in phase.
The maximum of $(\delta S)^2$ and work derivative coincide with the location of HBSh where on top of all $\xi(r)$ is large, and it is the 
combined constructive effect of all these facts which destabilizes \gov.
The $e-$folding time $\tau_e$ for \gov ~and other excited modes are less than a year.
This means that given an initial trigger, pulsational instabilities can develop fast (compared to the evolutionary time scales)
and saturate to reach an observable level on the surface.

On the right panel, \gst ~has large enough amplitude in the HBSh to provide enough energetics for destabilization, but the
radiative damping above the ICZ dissipates the mode energy so that the net cumulative work is negative, and this mode is
stable $\eta=-0.208$. Note different values of eigendisplacement between the right and left panels.
The Fe-bump of opacity embodies negligible amount of mass (see Section \ref{ss:interior}) and cannot inject enough energy to destabilize
\gst.

A noticeable difference between the aforementioned g-modes in Figure \ref{f:g-modes} is that the net kinetic energy $E_k$  
of the overstable mode is twelve orders of magnitude larger than that of the stable mode. 
The simple explanation for this is the larger amplitude of \gov ~at HBSh compared to \gst ~(see Eq. \ref{e:sigma_I}).
That is why those regions in the model 
that are capable of damping \gov ~have no effect on its net work $W$, while the radiative zones have enough energetics to dissipate 
the mechanical energy of \gst ~and damp it out.

\subsection{Any Hope on $\kappa-$Mechanism?}\label{ss:kappa}
The broad maxima of stable modes in Figure \ref{f:eta} correspond to gravity modes stabilized by the \kap ~arising from 
the bump of opacity associated with the He partial ionization zone at $\log T\approx4.6$. 
This zone is situated in the outer dilute envelope of Rigel and contains negligible mass (see the long-dashed curve in Figure \ref{f:interior}). 
As a result, the He partial ionization zone is incapable of attaining adequate inertia to drive gravity modes.

\begin{figure}[t]
\epsscale{1.15}\plotone{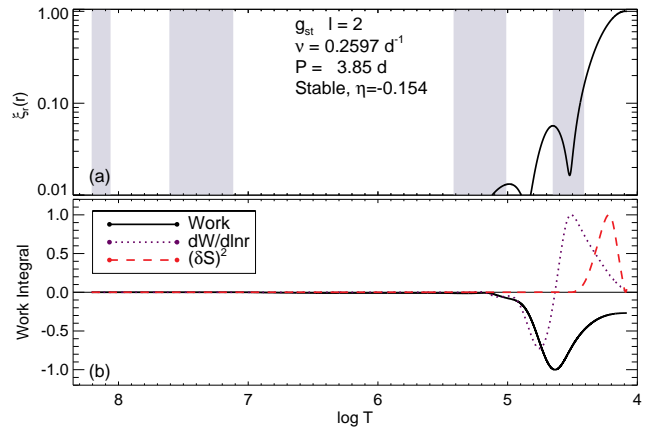}
\caption{An example of a quadrupole g-mode which is stabilized by \kap. 
The frequency and growth rate of this mode is selected from the broad peak in Figure \ref{f:eta}.
The Iron-bump has a slight stabilizing effect, but the He-bump contributes 
positively to the mode driving; the net effect, however, is mode stability.
\label{f:kap}}
\end{figure}

Figure \ref{f:kap} illustrates an $\ell=2$ g-mode selected from the broad maxima of stable modes depicted in Figure \ref{f:eta} 
with the frequency of $\sim$0.26 d$^{-1}$ and $\eta=-0.154$. 
The profile of $\xi_r(r)$ shown in Figure \ref{f:kap} panel (a) has a relatively small amplitude in the radiative zone between the 
Iron and He opacity bumps.
This is the reason for the negative work derivative $dW/d\ln r$ inward of the He bump (dotted line in Figure \ref{f:kap} panel b). 
However, there is still a positive and promising contribution to the work integral arising from the He bump as shown by the dotted line in panel (b).

% ------------------------------- Figure -------------------------------
% ------------------------------- Figure -------------------------------
% =================================================================================================================
% =============================================== CONCLUSION =====================================================
% =================================================================================================================

\section{Conclusion}\label{s:discuss}
The hydrogen burning shell in models of blue supergiant stars (like Rigel) is partly situated in the the radiative layer below the
intermediate convective zone. 
This is providing a situation where some modes can have a chance to overcome
the radiative damping by the \eps ~provided they have large amplitudes (or equivalently high kinetic energy $E_k$) in the radiative part of
the hydrogen burning shell.
This can occur when the heat exchange (perturbation to entropy change $(\delta S)^2$) reaches a maximum value at the H burning shell. 
When the location of maximum heat exchange between different stellar layers coincides spatially with the location of appropriate fluctuations 
of the generated nuclear energy, an instability is likely to take place.
Our current study which includes effects of rotational mixing can explain the long-term 
($21 \lesssim P \lesssim 127$ d) periodicities in Rigel.
The short-period limit (21 day) where \eps ~destabilizes some g-modes can possibly be altered by improvements in the physics of rotating stars
such as the efficiency of rotational mixing or the extent of convective overshooting from convective zones.

Furthermore, the close similarity between Rigel as a late B supergiant and other early A type supergiants in photometric and spectroscopic 
variability suggests extrapolating the application of the \eps ~to the long-term variation of the $\alpha$ Cyg class of pulsating stars.

The observed short-term spectroscopic variations of Rigel with periods below $\sim$10 days are still theoretically challenging 
\citep[see also][]{gautschy-2009}.
Previous works of \citet{dziembowski-1993-01, dziembowski-1993-02} and \citet{pamyatnykh-1999-01} showed that the Iron enhancement 
and the resulting increase in the height of opacity peak can help explain the overstability of main sequence pulsating B stars against 
heat driven p- and g-modes.
If the observed variations with periods shorter than about a week originate from Rigel's pulsations, not from other sources
(e.g. spots, variable winds, and propagating shocks), then \kap ~is the only plausible means to induce instabilities. 
Further local enhancement of Iron-group elements at $\log T\approx5.2$, in addition to He enhancement at $\log T\approx4.6$ 
can possibly lead to excitation of shorter period modes.
However, a more rigorous theoretical investigation of possible mechanisms to locally enhance the desired elements is called for. 

With the advent and success of the current space-based missions such as \textit{MOST}, \textit{CoRoT} and \textit{Kepler},
(almost) uninterrupted high precision photometry over long time baselines is made possible.
Supergiants such as Rigel and $\alpha$ Cyg type variable stars with possible long periods will greatly benefit from several months of continuous
multi-wavelength photometry as planned with BRITE nano-satellite constellation mission \citep{kuschnig-2009-01}.
The first of these nano-satellites is expected to be launched mid-2012. 
Their first observations are centered on Orion and should include Rigel in its field of view. 
This allows for the precise asteroseismology of blue supergiants (as progenitors of core collapse supernovae and their cores 
as seeds of the future neutron stars or black holes) to come to fruition, 
and bring a wealth of information about the near-core conditions of the pre collapse phase of the evolution of massive stars.

% =================================================================================================================
% ================================================= ENDINGS =======================================================
% =================================================================================================================
\acknowledgments 
\textbf{Acknowledgments} E.M. and E.G. dedicate this study to Professor Yousef Sobouti, the founder of Institute for Advanced Studies in Basic Sciences 
(IASBS), Zanjan, and the main contributor of modern astronomy and astrophysics in Iran.
We sincerely thank the anonymous referee for helping us improving the science content of the paper.
We appreciate the fruitful discussions with Dr. Bill Paxton (KITP, UCSB) for all of the assistance with the \texttt{MESA} code.
We are also grateful to comments from Dr. Conny Aerts (KULeuven) for reading the manuscript and for suggesting to use an input model 
that adequately resolves the hydrogen burning shell.
EM appreciates the warm hospitality and support he received from Observatoire Cote d'Azur and Villanova University during his visits.
A.M. acknowledges the funding of AstroMadrid (CAM S2009/ESP-1496) and
the Spanish grants ESP2007-65475-C02-02, AYA 2010-21161-C02-02.
This research was supported by NASA/MOST grant NNX09AAH28G which we gratefully acknowledge.

% =================================================================================================================
% ================================================= REFERENCES ===================================================
% =================================================================================================================
\bibliographystyle{apj}
\bibliography{apj-jour,../../my-bib}

\end{document}